\documentclass[twocolumn,showpacs,preprintnumbers,amsmath,amssymb]{revtex4-1}
\usepackage{graphicx}
\usepackage{dcolumn}
\usepackage{bm}
\usepackage{epsfig}
\usepackage{hyperref}
\begin{document}

\title{Unexpected decoupling of stretching and bending modes in protein gels}

\author{Thomas Gibaud$^{a*}$, Alessio Zaccone$^{b}$, Emanuela Del Gado$^{c}$, V\'eronique Trappe$^{d}$ and Peter Schurtenberger$^{e}$}

\affiliation{$^{a}$ Laboratoire de Physique, ENS Lyon, CNRS UMR 5672, 46 All\'{e}e d'Italie, 69007 Lyon, France}
\affiliation{$^{b}$Cavendish Laboratory, University of Cambridge, JJ Thomson Av.,
Cambridge CB3 0HE, U.K.}
\affiliation{$^{c}$ETH Z\"{u}rich, Institute of Building Materials, Z\"{urich}, Switzerland}
\affiliation{$^{d}$University of Fribourg, Department of Physics, Fribourg, Switzerland}
\affiliation{$^{e}$University of Lund, Division of Physical Chemistry, Lund, Sweden}

\date{\today}
\begin{abstract}
We show that gels formed by arrested spinodal decomposition of protein solutions exhibit
elastic properties in two distinct frequency domains, both elastic moduli exhibiting a remarkably strong dependence on volume fraction. Considering the large difference between
the protein size and the characteristic length of the network we model the gels as porous
media and show that the high and low frequency elastic moduli can be respectively attributed
to stretching and bending modes. The unexpected decoupling of the two modes in the
frequency domain is attributed to the length scale involved: while stretching mainly relates to
the relative displacement of two particles, bending involves the deformation of a strand with a
thickness of the order of a thousand particle diameters.
\end{abstract}
\pacs{83.80.Kn, 82.70.Gg. Published in: Phys. Rev. Lett. \textbf{110}, 058303 (2013).
} \maketitle

One of the overarching goals in soft condensed matter physics is to establish a rigorous connection between structural and mechanical properties of a material, and to relate those properties to control parameters such as the volume fraction, $\phi$, and the interaction strength, $U$, in a colloidal suspension. Particular attention has been given to the dynamical arrest transition from a fluid to a solid-like state \cite{2005cocis.trappe}. It has been shown that weakly attractive colloids ($U \sim k_{B}T$) at intermediate values of $\phi$ ($\sim 0.1 -0.4$) can form stress-bearing networks as a result of the interplay between spinodal decomposition and structural arrest \cite{pusey, verhaegh, Testard, 2005prl.manley, 2008n.lu, conrad, 2007prl.cardinaux, 2011sm.gibaud, 2007prl.buzzaccaro, 2011.sm.teecea, 2008.bj.dumetz, laurati}. This represents an unconventional route to gelation, where the arrest occurs when the dense domains formed during spinodal decomposition undergo structural arrest and phase separation can no longer proceed, the system being arrested as a self supporting network. In spite of the interest towards spinodal networks for many applications, there is still very little understanding of their mechanical behavior. An example of such systems is the solution of the globular protein lysozyme \cite{gogelein},
where the effective attraction between the proteins depends on temperature, which allows one to continuously explore the transition from homogeneous liquid to phase separated  and arrested states \cite{2009jpcm.gibaud}. This makes lysozyme a convenient system to investigate the interplay between structural evolution and arrest mechanisms. Previous studies showed that the gel states display correlation lengths much larger than the particle size \cite{2009jpcm.gibaud} but how such feature determines the mechanics remained unclear.

In this letter, we show that the rheological response of arrested lysozyme systems displays two distinct elastic plateaus that are well separated in time. Both moduli exhibit a dramatic dependence on volume fraction. Starting from these observations, we formulate a simple model where the two distinct moduli arise from stretching and bending modes of the coarse network structure, consisting of strands that are significantly bigger than the constituent particles. Our results provide guidance for the design of amorphous colloidal systems with elastic moduli that are remarkably sensitive to changes in $\phi$.

Our samples consist of hen egg white lysozyme (Fluka, L7651) with a radius $r\simeq1.7$~nm that are suspended in $20$~mM  HEPES buffer at pH =7.8, where lysozyme carries a net
charge of $+8e$. To screen the electrostatic repulsion we add salt, [NaCl]=500~mM
\cite{2007prl.cardinaux}. The total volume fraction $\phi$ is obtained from the protein
concentrations $c$ measured by UV absorption spectroscopy using $\phi=c/\rho$, where $\rho$=1.351~g/cm$^{3}$ is the protein density. As shown in previous work the phase behavior of our systems is determined by temperature \cite{2007prl.cardinaux,2009jpcm.gibaud}: the liquid-liquid phase boundary found at lower temperatures is metastable with respect to the liquid-crystal coexistence curve observed at higher temperatures. The spinodal decomposition is found to be arrested for quenches below the arrest tie line at $T$=15~$^{\circ}$C, which leads to a bicontinuous state with a protein-poor fluid network and a protein-dense glassy network, whose features depend on $\phi$ and $T$

We measure the rheological properties of such glassy networks with a commercial rheometer
using a cone and plate geometry where the temperature is controlled by a Peltier-element. In a typical experiment we load a sample in the rheometer at a temperature well above the binodal line at $T=25$~$^{\circ}$C. The temperature is then lowered to the final quench temperature at $T$=10~$^{\circ}C$, below the arrest tie line. The samples are left to rest for 300 s during which the structures stabilize \cite{2009jpcm.gibaud}. This procedure enables us to reach reproducible steady states with negligible aging. We explore different arrested states by varying the initial volume fraction $\phi$; any experiment where crystals form is disregarded.
 The viscoelastic properties of the arrested states are characterized in the linear response regime using two rheology experiments: (\emph{i})
oscillatory strain experiments yielding the elastic modulus $G'$ and the loss modulus $G''$ as a function of the frequency $f$, and (\emph{ii}) creep tests yielding the compliance $J$, as a function of time $t$, in response to a step stress.

The results of both experiments compare well with each other, as shown for $\phi$= 0.11 after a quench to $T=10$~$^{\circ}$C in Fig 1; the full squares and the open circles denote respectively $G'$ and $G''$ obtained in oscillatory strain experiment, while the lines correspond to $G'$ and $G''$ obtained by converting the data obtained in creep experiments \cite{2009pre.evans}. Strikingly, the mechanical response function of our system is characterized by two frequency domains where the elastic component $G'$ becomes dominant; these are marked by $G_{\infty}$ and $G_0$ at respectively the high and low frequency end of the spectrum shown in Fig. 1. Both characteristic elasticities are strong functions of the volume fraction, as shown in Fig. 2, where we display representative examples of the spectra obtained for different $\phi$ in creep experiments. Varying $\phi$ by only a factor of three leads to a variation of $G_{\infty}$ of four orders of magnitudes. To determine both characteristic elasticities with reasonable accuracy we use the scaling procedure described in EPAPS \cite{epaps} and report the results obtained in Fig. 3. Surprisingly, $G_{\infty}$ and $G_0$ increase nearly exponentially with $\phi$, where $G_{\infty}$ increases more quickly with $\phi$ than $G_0$, such that the two moduli appear to reach the same magnitude at large enough volume fractions.  Such exponential dependence is atypical for colloidal gels, where the $\phi$-dependence is generally well described by power-laws \cite{conrad, 2007prl.buzzaccaro, laurati, zukoski, Gruijthuijsen, trappe}. By contrast, stronger than power law dependencies of the elastic moduli on $\phi$ have been observed in ceramics and other porous
media \cite{pal}.

To rationalize our findings we thus consider the possibility to describe our systems as porous media. The structure of our systems has been probed in various scattering experiments \cite{faraday, 2009jpcm.gibaud}, from which we can infer that the arrested states are coarse network structures that are characterized by a correlation length of the order of $\xi \sim 2.5\mu m$. At somewhat smaller length-scales, larger scattering vectors $q$, we find the hallmarks of the Porod-regime, consistent with the gel being a bicontinuous network with a sharp interface between the dilute phase and the dense glassy phase. For $q \ge 0.15 nm^{-1}$, $S(q)$ follows a different regime, compatible with the presence of heterogeneities at small length-scales. Clearly, while the coarse network is characterized by a length scale that is $\sim$1000 times larger than the particle size the internal heterogeneities extend only over a few particle diameters. This large difference in size indicates that the system may be considered as almost homogeneous within the strand, being mainly characterized by large length scale heterogeneities, which is reminiscent of porous media.

\begin{figure}[h]
\begin{center}
\includegraphics[width=230pt]{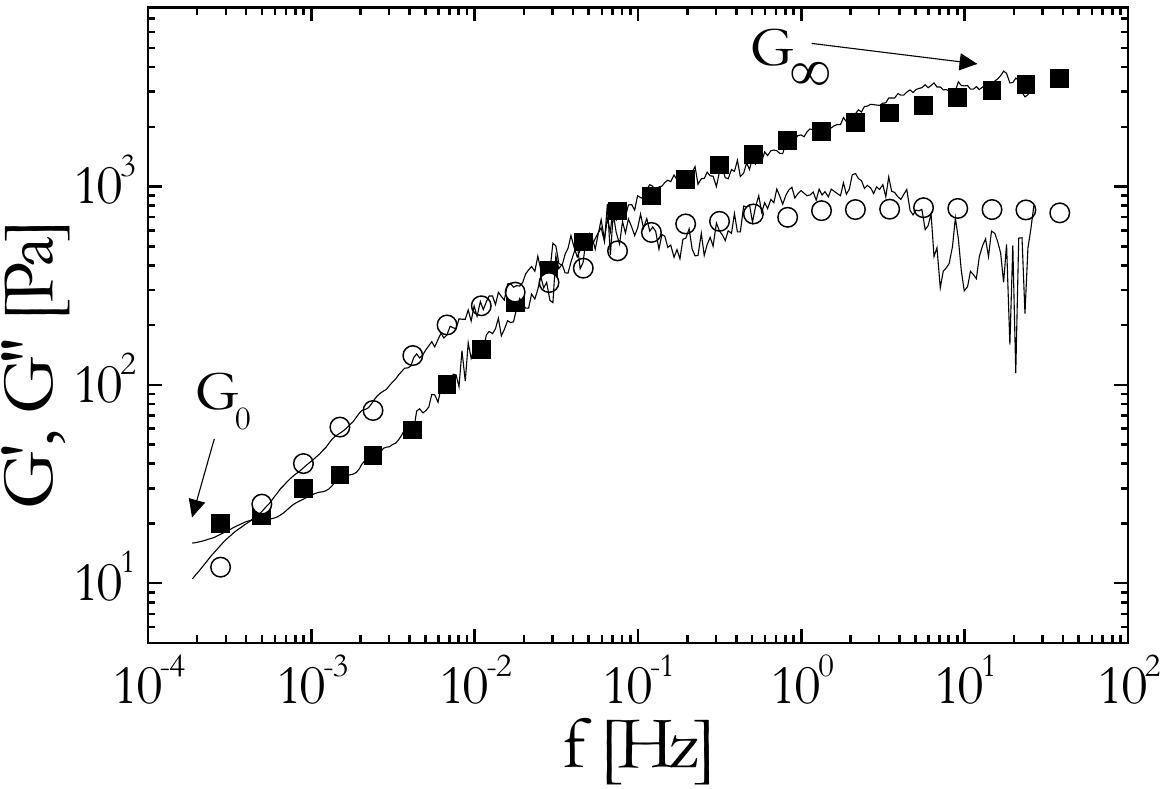}\\
\caption{Frequency dependence of the storage $G'$ and loss modulus $G''$ obtained for a lysozyme system with $\phi=0.11$ after a quench to $T=10$~$^{\circ}$C. Symbols denote the data obtained in oscillatory strain experiments, $G'$ ($\blacksquare$) and $G''$ ($\circ$). Lines correspond to $G'$ and $G''$ obtained by the conversion of the inverse compliance $J^{-1}$ measured in a creep experiments under identical conditions. $J^{-1}$ ($\blacktriangledown$) is displayed in Fig.~\ref{figure2}.}
\label{figure1}
\end{center}
\end{figure}

\begin{figure}[h]
\begin{center}
\includegraphics[width=220pt]{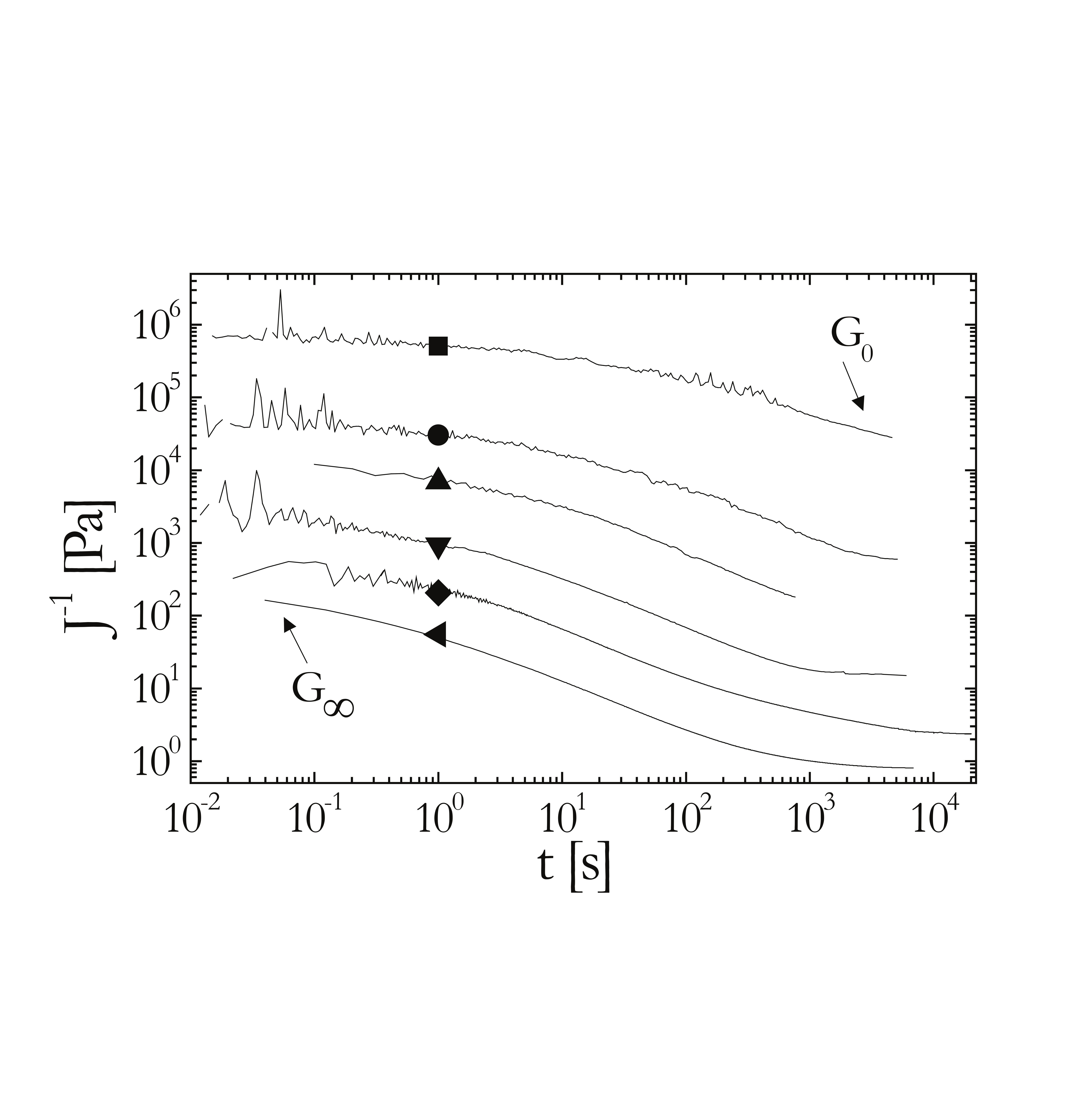}\\
\caption{Representive examples of the time evolution of the inverse compliance $J(t)^{-1}$  obtained in creep experiments using stresses that are well within the linear regime. The responses are measured after a quench to $T$=10~$^{\circ}$C for $\phi=0.171$ ($\blacksquare$), $0.149$ ($\bullet$), $0.13$ ($\blacktriangle$), $0.11$ ($\blacktriangledown$), $0.093$ ($\blacklozenge$) and $0.075$ ($\blacktriangleleft$).
}
\label{figure2}
\end{center}
\end{figure}

We then may conceive that the large mismatch between length scales is also at the origin of the unusual elastic response observed. Let us consider that the mechanical deformation results in an elastic response where both stretching and bending modes are involved \cite{frey}. In principle, the response should be dominated by the softer modes, which are the bending modes.  However, unbending a strand requires the cooperative displacement of several hundreds of particles, which may be delayed as compared to stretching modes which only involve the relative displacement of two particles. It is thus conceivable that we probe stretching modes at high $f$, while probing the bending modes at low $f$.

To test this assumption we consider that the stretching modulus $G_{s}$ is the main contribution to the high frequency response  $G_{\infty}$. We start from $G_{m}$ the shear modulus of a homogeneous assembly of proteins and estimate the correction due to the presence of pores. Given the lengthscale separation between the protein size ($nm$) and the strand size ($\mu m$) we can use an effective medium theory to account for the effect of the pores on the elastic properties \cite{torquato}. Starting from the stress contribution of a single inclusion to a homogeneous material, the effective medium theory allows us to calculate the elastic constants of a porous medium in analogy to Einstein's method for calculating the viscosity of a dilute suspension of spheres~\cite{einstein} and later extended by S. Arrhenius to higher $\phi$ \cite{arrhenius}. Because of the assumption of dilute and non-interacting pores, one first obtains a linear dependence of the properties of the porous material on those in absence of porosity: $G_{s}=G_{m}(1-g\epsilon)$, with $\epsilon$ the volume fraction occupied by the pores and $g$ a geometric pre-factor related in general to the pore geometry; it can vary from $g=5/3$ for spherical pores up to $~15$ for ellipsoidal pores upon increasing the ellipsoid's anisotropy ~\cite{pal,landau_FM}.
A small increment in the porosity leads to an increment of $G_{s}$: $dG_{s}=-g G_{s} d\epsilon$.  Upon integration with the condition $G_{s}=G_{m}$ at $\epsilon=0$, one obtains $G_{s} \simeq G_{m}  e^{-g\epsilon}$~\cite{pal}. The porosity $\epsilon$ is obtained from the volume fraction occupied by the dense network phase $h$, $\epsilon =1-h$. In our system $h$ increases linearly with $\phi$, as shown in Fig 4a where $h \simeq a \phi$ and $a\simeq 3.5$ \cite{2007prl.cardinaux}. We thus expect that $G_s$ scales as
\begin{equation}
G_{s} \simeq e^{g a \phi}
\label{poro}
\end{equation}

For the elastic response at the low-frequency end of the spectrum we assume that elasticity is dominated by
the strand elasticity and hence, because of the strand anisotropy, by the bending modes \cite{frey,scossa}. If we consider the strand as an elastic rod of length $\xi$ and diameter $m$ \cite{landau}, its bending energy is $K_{b}x^{2}$, where $x$ is the bending displacement and $K_{b}$ the bending constant. The bending energy can be also expressed in term of the bending moment $A$, i.e. the moment of forces due to internal stresses on a given cross-section of the rod. For small, pure bending deformations, $A = (E/R) (\pi/4) (m/2)^4$ ~\cite{landau}, where $E$ is the Young modulus and $R$ the curvature of the bent rod, that is $R \sim \xi$.
Moreover, by schematizing the strand as a chain of connected springs with an overall radius of gyration $\simeq \xi$, the bending displacement due to a force applied to the end of the chain is also of the order of $x\simeq\xi$ \cite{kantor}. As a consequence, the bending modulus is $G_{b} \simeq K_{b}/\xi \simeq A/\xi^{3} \simeq (m/\xi)^{4}E$. Using the shear modulus of the strand $G_{s}$ and eq.\ref{poro} one obtains
\begin{equation}
G_{b} \simeq (m/ \xi)^{4} G_{s} \simeq (m/ \xi)^{4} e^{g a \phi}.
\label{bending}
\end{equation}

For our system $\xi$ is almost independent of $\phi$, as shown in Fig 4b. However, since $h$ linearly depends on $\phi$ the strand thickness $m$ must depend on $\phi$. Considering that the volume fraction occupied by one strand is $V \simeq \xi m^2$ and assuming that the number density of strands is $1/\xi^{3}$, from the volume fraction occupied by the dense phase $h \simeq V/\xi^{3}$ we obtain the thickness of the network strands $m \simeq \sqrt{h}\xi$ \cite{conrad}, as reported in Fig. ~\ref{figure4}b.

\begin{figure}[h]
\centering
\includegraphics  [width=220pt] {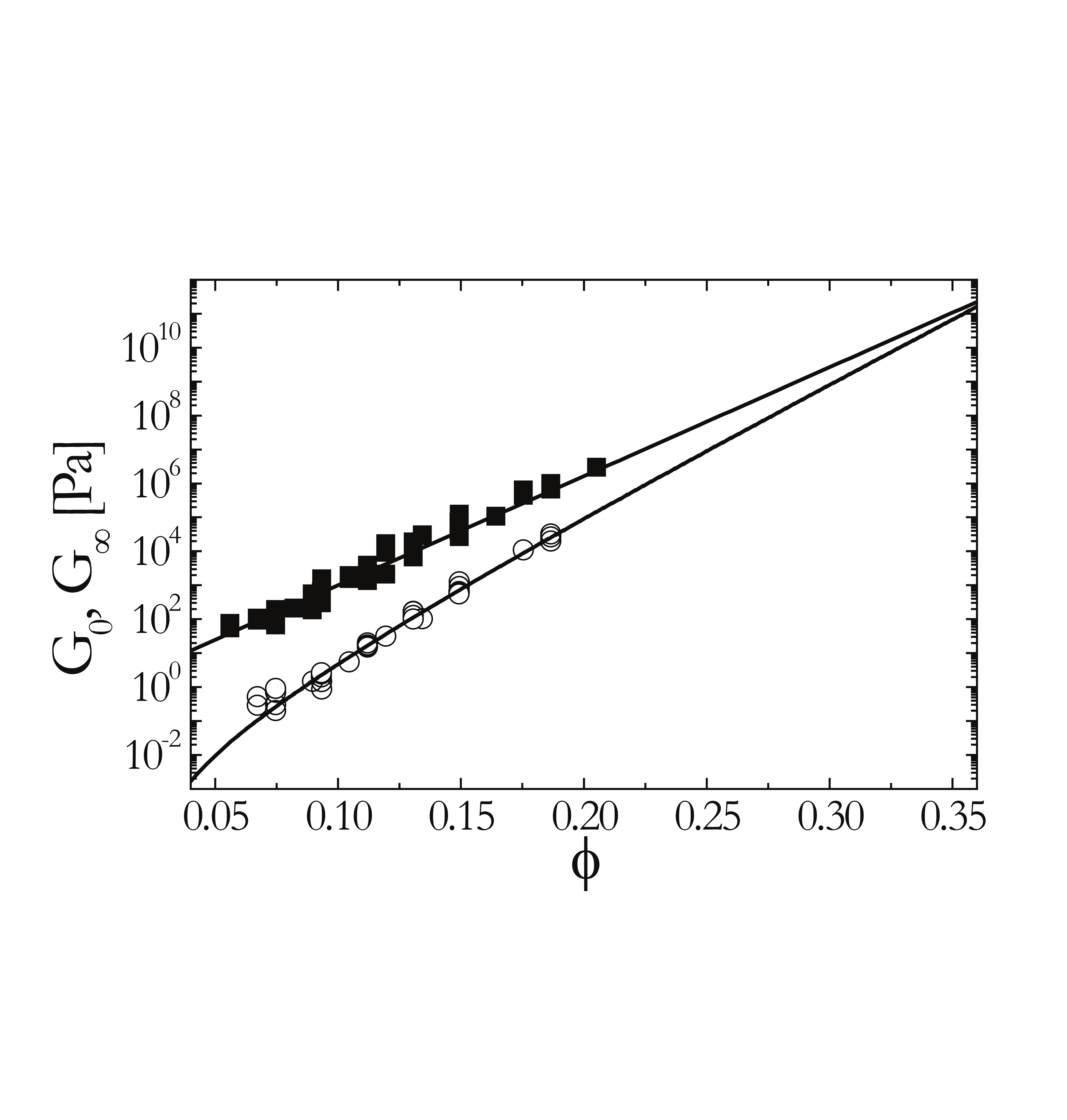}
\caption{Volume fraction dependence of high $G_{\infty}$ ($\blacksquare$) and low frequency modulus $G_{0}$ ($\bigcirc$). Lines are fits to the data using eq.\ref{poro} and \ref{bending} with $ga$=77$\pm$5.
}
\label{figure3}
\end{figure}

 \begin{figure}[h]
\centering
\includegraphics  [width=235pt] {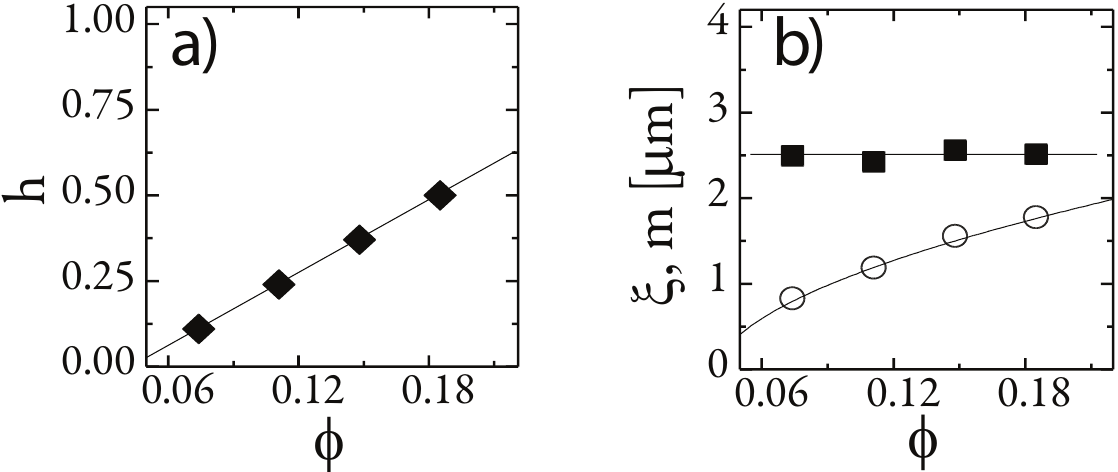}
\caption{Volume fraction dependence of the volume fraction occupied by the glassy network $h$ and the strand dimensions ($\xi$, $m$). a) $h$ increases linearly with $\phi$, $h=a\phi+b$ with $a$=3.5$\pm$0.3 and $b$=-0.15$\pm$0.03. b) $\xi$ ($\blacksquare$) is independent of $\phi$, while $m$ ($\bigcirc$) increases with $\phi$ as $m \sim \phi^{\alpha}$, with $\alpha$=0.55$\pm$0.05.
}
\label{figure4}
\end{figure}

With these experimental parameters we now can test the validity of our assumptions,  $G_{\infty}=G_s$ and $G_0 = G_b$. With $g=22.0\pm3.5$ we find that eq. 1 and 2 describe the volume fraction dependence of both moduli remarkably well, as shown in Fig 3 where the lines correspond to the theoretical prediction according eq 1 and 2. The remarkable agreement between prediction and experimental data supports the idea that $G_{\infty}$ is actually dominated by the contribution of the stretching modes, while $G_0$ is dominated by the bending modes  of the network strands. However, we note that the value we obtain for the parameter $g \simeq 22$ is larger than what is typically estimated in theory \cite{pal, landau_FM}. This large value for $g$ reflects the remarkable strong dependence of the moduli on $\phi$. Such increased dependence in comparison to that theoretically expected for porous media is most likely due to the structural differences between our very course networks and a porous medium comprising dilute and non-interacting pores, as assumed in our modeling. Moreover, we also neglected the heterogeneities inside the stands. Assuming that their contribution to the elastic modulus would linearly depend on $\phi$, this would lead to an enhanced dependence of $G_s$ on $\phi$ and to a value $g$ that is not only due to the pore, i.e. strands geometry.

Nonetheless, the good agreement between the mechanical properties of the arrested spinodal network and the scaling behavior predicted by our simple model supports the condition that stretching and bending modes are decoupled because of a delayed response of the bending mode. This feature is difficult to understand, but as denoted before could be related to the large differences in the length scales involved in the mescoscale organization of the network. Indeed, the bending of the thick strands may entail structural rearrangements of a large number of particles, which is a slow process. In contrast, stretching only involves local stretching of particle-particle bonds, which occurs instantaneously upon application of a stress.

Note, that the difference between particle and strand size also dictates the unusually strong dependence of the elastic modulus on volume fraction. Because of this difference the gels exhibit features of porous media, the elasticity increasing exponentially with $\phi$ instead of the typical power law increases observed in classical colloidal gels. Indeed, for colloidal systems, where spinodal decomposition is for instance induced by depletion \cite{conrad, laurati, zukoski, Gruijthuijsen,2008n.lu, 2005prl.manley, 2011.sm.teecea}, the arrest of the phase separation process generally leads to gels where the ratio between particle and strand size barely exceeds one order of magnitude. Both, bending and stretching modes should here still contribute to the elasticity of the system; however, it is unlikely that both modes will be as clearly separated in time, as this is the case in the lysozyme gels. For other protein-gels that are formed by spinodal decomposition, it appears likely that the mechanical response function will exhibit features similar to those observed here. Indeed, studies exploring the phase behavior of a series of different proteins \cite{2008.bj.dumetz} indicate that a large difference between protein and strand size may be a general feature of protein-systems that are deeply quenched into the liquid-liquid coexistence range. If this is indeed the case the resulting mechanical properties should be similar to those of the deeply quenched lysozyme systems. To make a definite assessment, however, further investigations of the volume fraction dependence of the structural and mechanical properties of deeply quenched protein systems are required.

In conclusion, we have investigated the rheology of the complex network structure that forms as a result of an arrested spinodal decomposition in solutions of the globular protein lysozyme. For these systems the characteristic correlation length of the network is found to be remarkably large as compared to the particle size. The mechanical response are characterized by two distinct moduli, $G_{0}$ and $G_{\infty}$ that are well separated in time, $G_{0}$ and $G_{\infty}$ being respectively associated to the long and short time response of the system. Using a porous medium approximation, we develop a simple model that links the structural features of the network to its rheological properties and their $\phi$-dependence. The remarkable agreement between theoretical prediction and experiment indicates that $G_{0}$ and $G_{\infty}$ are dominated by respectively the bending and the stretching modes of the arrested network. The two moduli are distinct and measurable because of the hierarchical mesoscale organization separating the network characteristic length ($\mu$m) and the protein size ($nm$):  this length-scale difference appears responsible for the separation of the time scales over which bending and stretching occurs. Our combination of experiments and modeling creates a framework for understanding the relation between rheology and structure. Our results may provide new guidelines to design amorphous solid-like network with specifically defined and controlled characteristic structural and mechanical properties with potential applications in materials and food science \cite{faraday, 2005nm.mezzenger, 2009sm.dufresne}.

\label{acknowledgments} This work was supported by the Swiss National
Science Foundation (grant No. PP002\_126483/1, 200020-117755 and 200021-127192),
the State Secretariat for Education and Research of Switzerland, the Marie Curie Network on Dynamical Arrest
of Soft Matter and Colloids (MRTN-CT-2003-504712) and the Agence Nationale de la Recherche fran\c{c}aise (ANR-11-PDOC-027).

* corresponding author: thomas.gibaud@ens-lyon.fr

\begin{center}
\textbf{EPAPS: determination of $G_0$ and $G_\infty$ using the scaling behavior of $J^{-1}$}
\end{center}

\begin{figure}[h]
\begin{center}
\includegraphics[width=220pt]{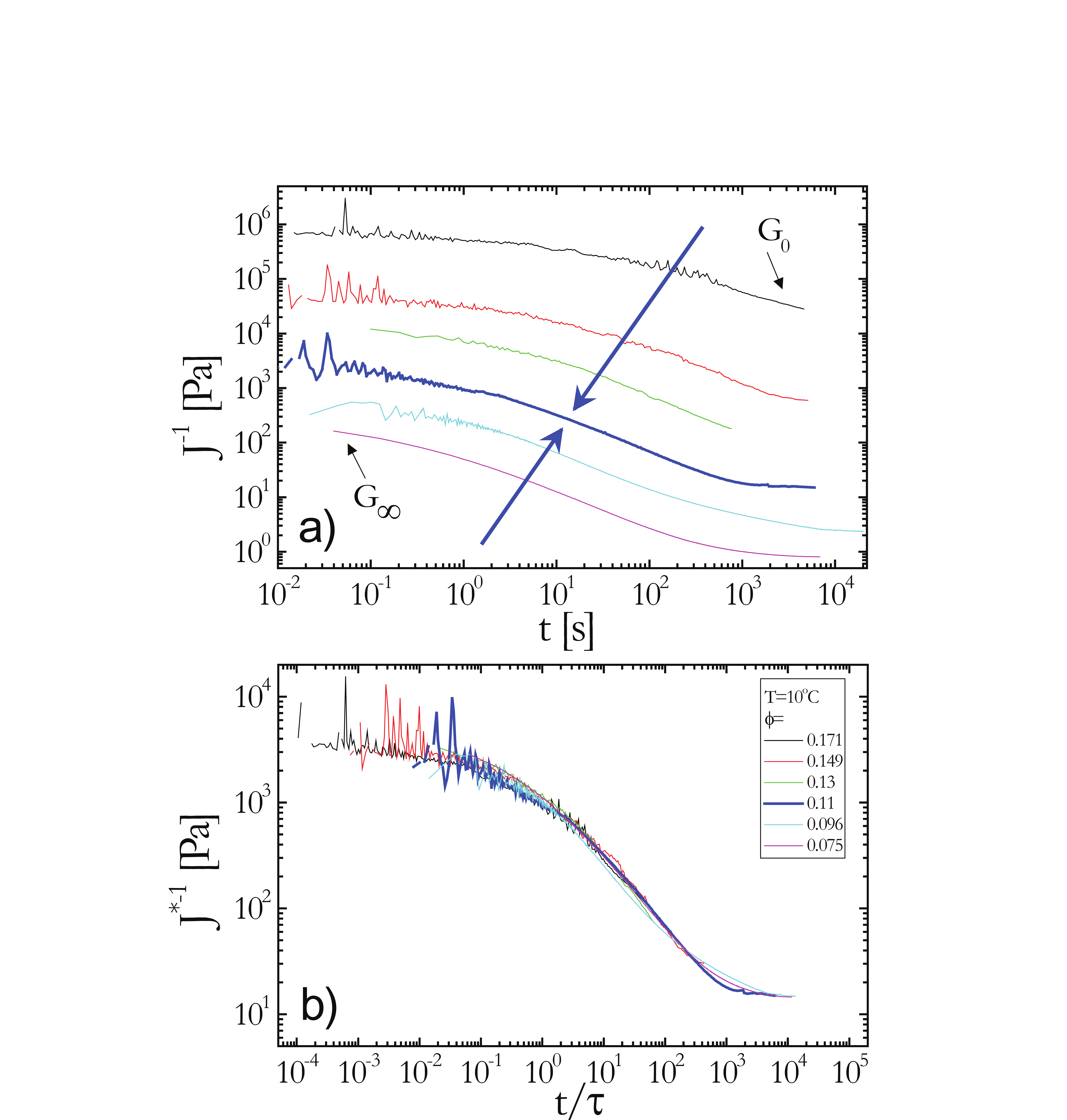}
\caption{Scaling of the inverse compliance $J^{-1}$ on a master curve $J^{*-1}$. a) Examples of the time evolution of the inverse compliance $J(t)^{-1}$ presented in Fig.2 in the article. From top to bottom: $T$=10~$^{\circ}$C, $\phi=0.171$, $0.149$, $0.13$, $0.11$, $0.093$ and $0.075$. $J^{-1}(\phi=0.11)$ is highlighted as it serve as the reference for the scaling procedure. b) Master curve,  $J^{*-1}$ based on the reference $J^{-1}(\phi=0.11)$, showing the scaled inverse compliance for the examples shown in a).}
\end{center}
\label{figure1}
\end{figure}
As shown in the EPAPS Fig.~1a (Fig. 2 in the article), the inverse compliance data not always extend to sufficiently long times to be able to properly determine the terminal modulus $G_0$, which should be determined in the time range where $J^{-1}$ become time independent. To also determine $G_0$ for the data-sets with a somewhat limited time window and to improve the relative accuracy of $G_0$ and $G_\infty$ obtained for the different samples, we use a scaling approach. In this scaling approach the functional form of the material response function of all our samples is presumed to be the same, differing only in the parameter $G_0$, $G_\infty$ and $\tau_c$,  where $\tau_c$ characterizes the time scale describing the transitional range between $G_\infty$ at short time scales and $G_0$ at long time scales. In praxis we fix the data set obtained with  $J^{-1}_{ref}(\phi)=J^{-1}(\phi=0.11)$ as reference and then scale all other data sets onto this curve by applying the following algorithm to $J^{-1}$:
\begin{align*}
J^{*-1}(\phi)&=J^{-1}_{ref}(\phi)\\&=\frac{J^{-1}(\phi)-G_{0}(\phi)}{G_{\infty}(\phi)-G_{0}(\phi)}[G_{\infty,ref}-G_{0,ref}]+G_{0,ref}.
\end{align*}
For the time axis we use $\tau=\tau_c/\tau_{ref}$. As shown in the EPAPS Fig. 1b, the scaling procedure works remarkably well supporting the hypothesis that the functional form of the material response function is the same for all our samples. We therefore use this scaling approach to determine $G_{0}$ and $G_{\infty}$ and the results are reported  in Fig.3 in the article.

\end{document}